\documentclass[pre,twocolumn,showpacs,floatfix,superscriptaddress]{revtex4}
 \usepackage{graphicx}
 
 \newcommand{\be}{\begin{equation}}
 \newcommand{\ee}{\end{equation}}
 \newcommand{\bea}{\begin{eqnarray}}
 \newcommand{\eea}{\end{eqnarray}}
 \newcommand{\xv}{\vec{x}}
 \newcommand{\ep}{\epsilon}
 \newcommand{\pb}{\bar{P}(m,t)}

\begin{document}

 \title{Persistence properties of a system of coagulating and annihilating
random walkers.}

 \author{Supriya Krishnamurthy}
 \email{supriya@santafe.edu}
 \affiliation{Santa Fe Institute, 1399 Hyde Park Road, Santa Fe, NM 87501,
USA}
 \author{R. Rajesh}
 \email{r.ravindran1@physics.ox.ac.uk}
 \affiliation{Department of Physics - Theoretical Physics, University of
Oxford, 1 Keble Road, Oxford OX1 3NP, UK}
 \author{Oleg Zaboronski}
 \email{olegz@maths.warwick.ac.uk}
 \affiliation{Mathematics Institute, University of Warwick, Gibbet Hill
Road, Coventry CV4 7AL, UK}


 \begin{abstract}
 We study a $d$-dimensional system
of diffusing particles that on contact either annihilate with probability
$1/(q-1)$ or coagulate with probability $(q-2)/(q-1)$. In $1$-dimension,
the system models the zero temperature Glauber dynamics of
domain walls in the $q$-state Potts model.
We calculate $\bar{P}(m,t)$, the probability that a randomly chosen lattice
site contains a particle whose ancestors have undergone 
exactly $(m-1)$ coagulations. Using perturbative renormalization
group analysis for $d < 2$, we show that, if the number of
coagulations $m$ is much less than
the typical number $M(t)$, then 
$\pb \sim m^{ \zeta/d} t^{-\theta}$, with
$\theta=d Q + Q( Q-1/2)  \epsilon + O(\epsilon^2)$, $\zeta=(2Q-1) \epsilon
+ (2 Q-1)  (Q-1)(1/2+A Q) \epsilon^2 +O(\ep^3)$, where $Q=\frac{q-1}{q}$,
$\epsilon =2-d$ and $A =-0.006\ldots$. $M(t)$ is shown to
scale as $M(t)\sim t^{d/2-\delta}$, where $\delta = d (1 -Q) + (Q-1)
(Q-1/2) \epsilon + O(\ep^2)$. In two dimensions, we show that $\pb \sim
\ln(t)^{Q(3-2Q)} \ln(m)^{(2Q-1)^2}t^{-2Q}$ for $m \ll t^{2 Q-1}$.  The
$1$-dimensional results corresponding to $\epsilon=1$ are compared with
results from Monte Carlo simulations. 
 \end{abstract}
 \maketitle

\section{\label{sec1}Introduction}

Persistence is understood as a property of an evolving system to
``remember'' its initial configuration for anomalously long times. A
particular case of persistence that has received much attention is that of
site persistence (see \cite{SM} for a review).  The site persistence
probability is defined as the probability that the values of a
dynamical variable at a given set of sites do not change up to time $t$. 
For instance, in a spin system this could be the probability that a spin
at a given site does not flip up to time $t$, or in a reaction-diffusion
system, the probability that no reaction takes place at that site up to
time $t$.  In many cases the site persistence probability decays at large
times as a power-law \cite{cardy1}.

A natural generalization of site persistence is persistence of a pattern
present in the initial configuration \cite{MC}.  An instance of pattern
persistence would be the survival of a test particle in a random
environment. Examples of the random environment include diffusing traps
\cite{DV,MG}, reaction-diffusion systems such as $A\emptyset
\leftrightarrow \emptyset A$ \cite{DB1}, $A+B \rightarrow \emptyset$
\cite{DB1,BB} or $A_i + A_j \rightarrow A_{i+j}$ with mass dependent
diffusion rates \cite{alava1,alava2,KRZ1}, and predators in predator-prey
models \cite{KR1,CK}.  The 
problem of calculating the survival probability of the test particle is hard,
mainly because in the rest frame of the test particle,
the motion of the other particles is correlated.

The $1$-dimensional Potts model has been a testing ground for various
concepts of persistence.  The site persistence problem mentioned above, 
has been exactly solved for the $1$-dimensional $q$-state
Potts model evolving via zero temperature Glauber dynamics \cite{derrida}. 
Besides site persistence,
several other persistence properties of the Potts model have been studied. 
Among these are the probability that a domain wall has never encountered
another domain wall \cite{monthus,MC,DB2}, and the probability 
that a domain present in the initial configuration survives up to time $t$ \cite{NK}. 
The former problem has been studied numerically
\cite{MC,DB2}, by mean field approximations \cite{MC} and perturbatively
near $q=1$ \cite{monthus}. 
However, the results obtained by these
techniques
do not approximate well the numerical results in the whole range
of $q$.

In dimensions greater than one, the dynamics of domain walls in the
Potts
model is difficult to treat analytically. Instead, we note that in
$1$-dimension, the zero temperature Glauber dynamics of the $q$-state
Potts model is equivalent to a system of diffusing particles that on
contact either annihilate with probability $1/(q-1)$ or coagulate with
probability $(q-2)/(q-1)$ \cite{D,satya,SC}. We study the persistence
properties of this reaction-diffusion system in an arbitrary number of
dimensions using the renormalization group method and calculate the
exponents as an $\epsilon$-expansion. 

The question that we ask is, given this reaction-diffusion system, what is
the fraction of particles that have never encountered another particle up
to time $t$? More generally, what is the fraction of particles whose
ancestors have
undergone $m$ coagulations up to time $t$?  A convenient way to keep track
of the history of coagulations is to assign a mass to each particle as
follows.  At time $t=0$, let all particles be of mass one.  Each time two
particles coagulate, the
new particle has a mass which is the sum of the
masses of the two parent particles.  It is clear that the particles
of mass $m$ will be those whose ancestors have undergone exactly
$(m-1)$ coagulations.

Let $\bar{P}(m,t)$ be the probability that a randomly chosen site at time
$t$ contains a particle of mass $m$. Let $\bar{N}(t)$ and $\bar{\rho}(t)$
denote the average particle density and average mass density respectively.
Then, the probability distribution $\bar{P}(m,t)$ is expected to have the
scaling form
 \be
 \bar{P}(m,t)=\frac{\bar{N}(t)^2}{\bar{\rho}(t)} f
\left(\frac{l_m}{\sqrt{t}}\right), \label{eq:2}
 \ee
 where $l_m = (m/\bar{\rho}(t))^{1/d}$ is the length scale associated with
mass. The large time behavior of $\bar{\rho}(t)$ and $\bar{P}(m,t)$ is
characterized by two exponents $\delta$ and $\theta$. The mass
density $\bar{\rho}(t) \sim t^{-\delta}$.  For masses much smaller than the
typical mass, $\bar{P}(m,t)$ decays as a power law in time as $m^{\zeta/d}
t^{-\theta}$. We will call $\theta$ the persistence exponent. The
exponent $\zeta$ characterizes the small $x$ behavior of the scaling
function $f(x)$ to be $f(x) \sim x^\zeta$ for $x \ll 1$. Then, $\zeta = 2
d (\theta+\delta-d)/(d-2 \delta)$, where we have used the fact that $\bar{N}(t)
\sim t^{-d/2}$ in dimensions less than two \cite{Peliti}.  The two
independent exponents $\theta$ and $\delta$ are known for some limiting
cases. 

When $q=2$, the model reduces to the reaction diffusion model $A+A
\rightarrow \emptyset$. All particles are of mass one and it is known that
$\bar{P}(1,t)  \sim t^{-d/2}$ for $d<2$ and $\bar{P}(1,t) \sim \ln(t)/t$
for $d=2$ \cite{aa,Peliti}. Hence, $\delta=d/2$, $\theta=d/2$ and
$\zeta=0$ for $q=2$. When $q=\infty$, the model is equivalent to the
reaction-diffusion system $A_i+ A_j \rightarrow A_{i+j}$
\cite{spouge,bram,KR,satya,oleg,KRZ1}. Since mass is conserved, $\delta=0$. It
has been shown that $\theta = d+\epsilon/2 + O(\epsilon^2)$, and
$\zeta=\epsilon + O(\epsilon^2)$ for $\epsilon=2-d>0$ \cite{KRZ1}.  In
$2$-dimensions $\bar{P}(m,t) \sim \ln(m) \ln(t)/t$ \cite{KRZ1}. In
$1$-dimension, it is known via an exact calculation that $\bar{P}(m,t)
\sim m t^{-3/2}$ \cite{spouge}. When $q\approx 1$, $\theta$ has been
calculated perturbatively to be $\theta= (q-1) 3 \sqrt{3}/(2 \pi) +
O((q-1)^2)$ \cite{monthus}. However for arbitrary values of $q$, the only
known analytical result follows from a mean field approximation \cite{MC}.
But the numerically obtained value for $\theta$ is very different
from the mean field value.  In this paper, we address this issue by using
the renormalization group formalism to systematically calculate
$\bar{P}(m,t)$ for arbitrary $q$. 

We now summarize our main results and give an outline of the rest of the
paper. In Sec.~\ref{sec2}, we give a precise definition of the model
and derive the stochastic partial differential equations obeyed by the
mass distribution.  In Sec.~\ref{sec3} we express the exponent $\delta$ in
terms of the exponent $\theta$, though at a different value of $q$,
reducing the number of unknown exponents to one. We show that
 \be
 \delta(q) = \theta\left(\frac{q}{q-1}\right).  \label{beta1}
 \ee
 Thus,
 \be
 \zeta(q) = \frac{2 d \left[\theta(q) +\theta\left(\frac{q}{q-1}\right) 
-d \right]} {d - 2 \theta\left(\frac{q}{q-1}\right)}. \label{gamma1}
 \ee

In Sec.~\ref{sec4} we use the technique developed in \cite{KRZ1} to
calculate the persistence exponent $\theta$ as an $\epsilon$-expansion,
where $\epsilon = 2-d>0$. We show that
 \be
 \theta = d Q + Q (Q-\frac{1}{2}) \epsilon + O(\epsilon^2),
 \label{eq:1}
 \ee
 where $Q=\frac{q-1}{q}$.  If $d=2$, the scaling form Eq.~(\ref{eq:2})  breaks
down due to logarithmic corrections. We calculate these corrections to be
 \be
 \pb \sim \frac{\ln(t)^{Q(3-2Q)}\ln(m)^{(2Q-1)^2}}{t^{2Q}},\label{pmt2d}
 \ee
 given that $t\rightarrow \infty$ and $m \ll M(t)$, where $M(t)$ is mass
of a typical particle at time $t$.
The analytical results for $\theta$ and $\delta$ in $1$-dimension obtained by
putting $\epsilon=1$ are compared with the results from numerical
simulations.

In Sec.~\ref{sec5} we show that the coefficient of $\epsilon^n$ in
Eq.~(\ref{eq:1}) is a polynomial of degree $2n$ in the variable
$Q=(q-1)/q$. This observation allows us to calculate the $2$-loop
corrections to the exponent $\zeta$ to be
 \be
 \zeta=(2 Q-1) \epsilon + (2 Q-1) (Q-1) (\frac{1}{2}+ A Q)
\epsilon^2+O(\ep^3),
 \ee
 where $A= -0.006\ldots$. 
The analytical results for $\zeta$ in $1$-dimension obtained by
putting $\epsilon=1$ are compared with the results from numerical
simulations.

Finally, we conclude with a summary and discussion in Sec.~\ref{sec6}.

\section{\label{sec2}Model and field theoretic formulation}

In this section, we define the model and derive the stochastic partial
differential equation obeyed by the mass distribution.  Consider a
$d$-dimensional lattice whose sites may be occupied by particles
that possess a positive integral mass. Multiple occupancy of a lattice
site is allowed. Given a certain configuration of particles on this
lattice, the system evolves in time via the following microscopic moves.
(i) With rate $D$, each particle hops to a nearest neighbor lattice site.
(ii) With rate $\lambda_c$, two particles at the same site coagulates
together to form a new particle whose mass is the sum of the masses of the
two parent particles. (iii) With rate $\lambda_a$, two particles at
the same site annihilate each other. To make connection with the model
discussed in the introduction, we have to choose $\lambda_c = \lambda
(q-2)/(q-1)$ and $\lambda_a=\lambda/(q-1)$ where $\lambda$ is a reaction
rate. The limit $\lambda \rightarrow \infty$ corresponds to instantaneous
reactions. In dimensions $d\leq 2$ and in the limit of large time, the
statistical properties of a finite reaction rate particle system were
shown to be equivalent to those of a system with infinite reaction rates
\cite{KRZ1}.  However, from the field theoretic point of view, it is more
convenient to work with finite reaction rates, and hence $\lambda$ will be
taken to be finite in this paper. 

Starting from the master equation for the time evolution of the system, we
now derive the effective field theory of the model.  Let $\{n_i\}$ denote
the configuration of particles at site $i$ such that $n_{i,m}$ is the
number of particles of mass $m$ at site $i$.  Let ${\mathcal{P}}(\ldots
\{n_i\}, \{n_j\}, \ldots;t)$ be the probability of the configuration
$(\ldots \{n_i\}, \{n_j\}, \ldots)$ at time $t$, where $i$ and $j$ are
nearest neighbors.  The master equation describing the time evolution of
${\mathcal{P}}(\ldots \{n_i\}, \{n_j\}, \ldots;t)$ is 
 \begin{widetext}
 \bea \lefteqn{\frac{d {\mathcal{P}}(\ldots \{n_i\}, \{n_j\}, \ldots)}{dt}
= - D \sum_{\langle i j \rangle} \bigg[ \sum_{m} (n_{i,m}+ n_{j,m})
{\mathcal{P}}(\{n_i\}, \{n_j\})  - \sum_{m} (n_{i,m}+1)
{\mathcal{P}}(\{n_{i,m}+1\}, \{n_{j,m}-1\}) } \nonumber \\ &&\mbox{} -
\sum_{m} (n_{j,m}+1) {\mathcal{P}}(\{n_{i,m}-1\}, \{n_{j,m}+1\}) \bigg]
-\lambda_c \sum_{i} \bigg[ \sum_{m\neq m'} n_{i,m} n_{i,m'}
{\mathcal{P}}(\{n_i\}) + \sum_{m} n_{i,m} (n_{i,m} -1 )
{\mathcal{P}}(\{n_i\}) \nonumber\\ && \mbox{} - \!\!\! \sum_{m\neq m'}
\!\!\!  (n_{i,m}+1) ( n_{i,m'}+1) {\mathcal{P}}(\{n_{i,m}+1, n_{i,m'}+1,
n_{i,m+m'}\!- \!1\})  - \sum_{m} (n_{i,m}+2) ( n_{i,m}+1)
{\mathcal{P}}(\{n_{i,m}+2, n_{i, 2 m}\!-\!1\})  \bigg] \nonumber \\
&&\mbox{} - \lambda_a \sum_{i} \bigg[ \sum_{m\neq m'} n_{i,m} n_{i,m'}
{\mathcal{P}}(\{n_i\}) + \sum_{m} n_{i,m} (n_{i,m} -1 )
{\mathcal{P}}(\{n_i\})  - \!\!\! \sum_{m\neq m'} \!\!\! (n_{i,m}+1) (
n_{i,m'}+1) {\mathcal{P}}(\{n_{i,m}+1, n_{i,m'}+1\}) \nonumber \\ &&
\mbox{} - \sum_{m} (n_{i,m}+2) ( n_{i,m}+1) {\mathcal{P}}(\{n_{i,m}+2\})
\bigg],
 \label{master}
 \eea
 \end{widetext}
 where the time dependence of ${\mathcal{P}}$ has been dropped for
notational simplicity and $\{n_{i,m}+1\}$ denotes the configuration
$(n_{i,1},n_{i,2}, \ldots, n_{i,m}+1, \ldots)$ at site $i$.  The first
term in the right hand side of Eq.~(\ref{master})  describes the loss and
gain terms arising from particles diffusing to their nearest neighbors
with rate $D$.  The second term describes the loss and gain terms 
due to the coagulation of a pair of
particles at a site with rate $\lambda_c$ to form a new particle whose
mass is the sum of the constituents. The third term describes the
loss and gain terms due to 
annihilation of a pair of particles at a site with rate $\lambda_a$. 

The field theory corresponding to the problem can be derived from the
master equation using Doi's formalism \cite{Doi}. In short,
regarding the master equation as a Schroedinger equation in imaginary
time, the functional integral representation of the corresponding
non-Hermitian evolution operator is constructed. This allows one to
write
down a functional integral expression for any correlation function of
the
problem, including $\pb$.  After taking the continuum limit, one is left
with the problem of solving an interacting field theory. The application
of Hubbard-Stratonovich transformation to this field theory leaves one
with a stochastic partial differential equation.  We refer to
Refs.~\cite{cardy,oleg,mattis} for reviews of this procedure.  Following
this procedure, solving the master equation Eq.~(\ref{master}) is
equivalent to solving the following Langevin equation for a stochastic
field $\tilde{P}(\xv,m,t)$:
 \bea
 \lefteqn{(\partial_{t}-D\nabla^2 )\tilde{P}(\xv, m, t) =} \nonumber \\
 &&\mbox{}- 2(\lambda_{c} +\lambda_{a})\tilde{P}(\xv, m,t) 
\int_0^{\infty} dm' \tilde{P}(\xv,m',t) + \lambda_{c} \tilde{P}*\tilde{P}
\nonumber \\
 &&\mbox{} + i\sqrt{2(\lambda_{a}+\lambda_{c})}\xi (\xv,t) \tilde{P}(\xv,
m, t), \label{sse}
 \eea
 where $\tilde{P}*\tilde{P}=\int_{0}^{m} dm'\tilde{P}(\xv,m', t) 
\tilde{P}(\xv, m-m', t)$, $\xi$ is white noise in space and time with unit
standard deviation and $i^2=-1$. 

The stochastic field $\tilde{P}(\xv,m,t)$ is complex and is different from
the local mass distribution $P(\xv,m,t)$, which denotes the number of
particles of mass $m$ in the volume $d^dx dm$ at time t. However, the moments
of $P$ are related to the moments of $\tilde{P}$ (for instance, see
\cite{mattis,Lee}).  For example, $\overline{P(\xv, m, t)}=
\overline{\tilde{P} (\xv, m, t)}$, $\overline{ P(\xv, m, t)^2 }=
\overline{ \tilde{P} (\xv, m, t)} (\Delta m (\Delta x)^d)^{-1} +
\overline{ \tilde{P}(\xv, m, t)^2 }$, and so on, where the bar on top
denotes an averaging over noise, and $\Delta x$ and $\Delta m$ are lattice
cutoffs.  In this paper we only study the first moment of $P(\xv, m, t)$,
and hence disregard the difference between $P(\xv, m, t)$ and
$\tilde{P}(\xv,m,t)$ in the rest of the paper. 

We will be studying the behavior of the following three quantities,
 \bea
 P(m,\xv,t) &\mbox{for }& m \ll M(t), \\
 N(\xv,t) &=& \int_{0}^{\infty} dm P(m,\xv,t),\\
 \rho(\xv,t) &=& \int_{0}^{\infty} dm~ m P(m,\xv,t),
 \eea
 where $N(\xv,t)$ is the local particle density, $\rho(\xv,t)$ is the
local mass density and $M(t) \sim \bar{\rho} (t)/\bar{N}(t)$ is the
typical mass at time $t$.  The time evolution equations obeyed by
$N(\xv,t)$ and $\rho (\xv,t)$ are easily obtained from Eq.~(\ref{sse}). As
for $P(m,\xv,t)$, for $m \ll M(t)$, we neglect the convolution term in the
right hand side of Eq.~(\ref{sse}). This approximation is justified
because at large times the probability of collision of two light particles
is negligible compared to the probability of collision of light and heavy
particles. The resulting equations are: 
 \bea
 \lefteqn{(\partial_{t}-D\nabla^2 )N(\xv, t)= -
(\lambda_{c}+2\lambda_{a})N(\xv, t)^2 } \qquad \qquad \nonumber \\ &&
\mbox{}
 + i\sqrt{2(\lambda_{a}+\lambda_{c})}\xi (\xv,t) N(\xv,
 t), \label{Nevol}
 \eea
 \bea
 \lefteqn{\!\!\!\!(\partial_{t}-D\nabla^2 )P(\xv,m, t)= -2
(\lambda_{c}+\lambda_{a}) P(\xv,m, t)N(\xv, t)  } \qquad \qquad \qquad
\quad \nonumber\\ && \mbox{} \!\!\!\!\!\!\!  +
i\sqrt{2(\lambda_{c}+\lambda_{a})}\xi (\xv,t) P(\xv,m, t), \label{ssep0}
  \eea
  \bea
 \lefteqn{(\partial_{t}-D\nabla^2 )\rho (\xv, t)  = -2 \lambda_{a} \rho
(\xv, t)N(\xv,t)} \qquad \qquad \nonumber\\
 &&\mbox{}
  + i\sqrt{2(\lambda_{c}+\lambda_{a})}\xi (\xv,t) \rho(\xv, t).
\label{sser0}
 \eea
 Note that the dependence of $P$ on mass is no longer governed by
Eq.~(\ref{ssep0}). Once the time dependence of $\bar{P}$ is calculated,
its mass dependence can be restored using dimensional analysis. In the
rest of the paper, for the sake of notational simplicity, we omit the
dependence of $P$ on mass, unless there is a cause for confusion. 

Equations~(\ref{Nevol}), (\ref{ssep0}) and (\ref{sser0}) can be simplified
as follows. Let
 \bea
 \lambda_c &=& \frac{q-2}{q-1} \lambda, \\
 \lambda_a &=& \frac{1}{q-1} \lambda
 \eea
 for some parameter $\lambda$.  Rescaling the local particle density,
local mass distribution and average density according to
 \bea
 \lefteqn{
 \bigg( N(\xv,t), P(\xv,t),\rho (\xv,t) \bigg) 
 \longrightarrow} \qquad \nonumber \\
 && \left( \frac{q-1}{q} \right) \bigg( N(\xv,t), P(\xv,t), \rho (\xv,t) 
\bigg),
 \eea
 brings Eqs.~(\ref{Nevol})--(\ref{sser0}) into the following form: 
 \bea
 (\partial_{t}-D\nabla^2 )N(\xv, t)&=&- \lambda N^2 (\xv, t)\nonumber\\
 && \mbox{} + i\sqrt{2\lambda}\xi (\xv,t) N(\xv, t), \label{ssen}\\
(\partial_{t}-D\nabla^2 )P (\xv, t)&=& -2 Q \lambda P (\xv, t)N(\xv,t) 
\nonumber\\
 && \mbox{} + i\sqrt{2\lambda}\xi (\xv,t) P(\xv, t),
 \label{ssep}\\ (\partial_{t}-D\nabla^2 )\rho (\xv, t)&=& -2 (1-Q) \lambda
\rho (\xv, t) N(\xv,t)  \nonumber\\
 && \mbox{} + i\sqrt{2\lambda}\xi (\xv,t) \rho(\xv, t), \label{sser}
 \eea
 where
 \be
 Q = \frac{q-1}{q}. 
 \ee
 It can be shown that Eqs.~(\ref{ssen}) and (\ref{ssep})  describe the two
species reaction $A+A \stackrel{\lambda}{\rightarrow}A$,
$A+B\stackrel{Q\lambda}{\rightarrow}\emptyset$, in the limit when the
concentration of $A$-particles is much greater than concentration of
$B$-particles. 

\section{\label{sec3}Scaling analysis of stochastic evolution equations.}

In this section, we obtain some exact results for the model.  First, the
scaled density of particles $N(\xv,t)$ obeys the same equation as the
particle density in the $A+A \rightarrow A$ reaction. For this reaction,
the density of particles decays for large times as $t^{-d/2}$ in $d<2$
\cite{aa,Peliti}.  Thus,
 \be
 N(t) = c \frac{q-1}{q}\frac{1}{t^{d/2}}, ~ t \rightarrow \infty, 
 \label{eq:exact}
 \ee
 where $c$ is a constant depending on dimension only. This is a
generalization of the exact $1$-dimensional result 
\cite{zeitak,masser}.

Secondly, there is a relation between the local mass distribution of light
particles $P$ and the local mass density $\rho$. Under the substitution $Q
\rightarrow (1-Q)$, or equivalently $q/(q-1) \rightarrow q$,
Eq.~(\ref{ssep}) transforms into Eq.~(\ref{sser}).  Therefore, if
$F_{P}(Q, \xv_{1},t_{1}, \xv_{2}, t_{2}, \ldots )$ is a correlation
function of $P$-fields, which is independent of initial conditions, then
$F_{P} (1-Q, \xv_{1}, t_{1}, \xv_{2}, t_{2}, \ldots)$ is the correlation
function of the same configuration of $\rho$-fields. In particular, since
$\bar{P}(t) \sim t^{-\theta(Q)}$, we obtain $\rho(t) \sim
t^{-\theta(1-Q)}$.  Thus, we derive Eq.~(\ref{beta1}), namely
 \be
 \delta(Q) = \theta(1-Q). 
 \label{relation2}
 \ee

We now examine Eqs.~(\ref{ssen}) and (\ref{ssep}) for special values of
the parameter $Q$. When $Q=0$, the non-linear term in Eq.~(\ref{ssep}) 
vanishes and the concentration of monomers is conserved on average.
Therefore, $\theta=0$ for $Q=0$. When $Q=1/2$, Eq.~(\ref{ssep}) is solved
by $P \sim N$, where $N$ is a solution of Eq.~(\ref{ssen}). Then, from
Eq.~(\ref{eq:exact}), we obtain $\theta=d/2$ for $Q=1/2$. When $Q=1$, it
is known that $\theta = d +\epsilon/2+O(\epsilon^2)$, where $\epsilon=2-d$
\cite{KRZ1}. If $d=1$, then $\theta=3/2$, which is a consequence of an
exact solution \cite{spouge}, rather than the $\epsilon$-expansion cited
above.  Collecting these results together, we have
 \bea
 \theta = \cases{
 0 & for $Q = 0$, \cr
 \frac{d}{2} & for $Q = \frac{1}{2}$, \cr
 d + \frac{\epsilon}{2}+O(\epsilon^2) & for $Q = 1$. \cr}
 \label{exact}
 \eea

The question remains as to whether $\theta$ for $Q<1/2$ or equivalently
$q<2$ has any physical implication. For the site persistence problem in
$1$-dimension, it is known that the site persistence probability of the
$q$-state Potts model maps to an Ising system with an initial
magnetization given by $2/q -1$, evolving via zero-temperature Kawasaki
dynamics \cite {satya,SC,BFK}.  The latter system is defined at any value
of $q >1$. The correspondence between these two models also holds for the
persistence probability of a single domain in the Potts model \cite{NK}.
It would be interesting to understand what quantity, if any, in the Ising
model corresponds to the survival probability of domain walls in the Potts
model.

\section{\label{sec4}Perturbative computation of persistence exponent near
$d=2$.}

In this section, we calculate the large time behavior of $\bar{P}(t)$
using the formalism of perturbative renormalization group. We closely
follow the solution of the $A_{i} + A_{j} \rightarrow A_{i+j}$ model
presented in Ref.~\cite{KRZ1} . 

\begin{figure}
 \includegraphics[width=\columnwidth]{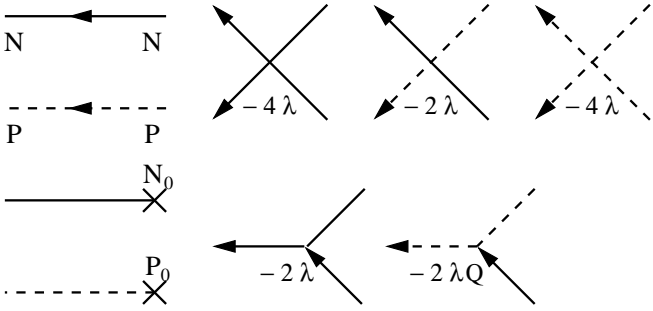}
 \caption{\label{feynmanfig1} Propagators and vertices of the theory.}
 \end{figure}

The solution to $\bar{P}(t)$ as a perturbative expansion in powers of
$\lambda$ can be constructed from Eqs.~(\ref{ssen}) and (\ref{ssep}) using
Feynman diagrams \cite{Drouffe}. The Feynman rules for constructing
terms of the expansion are summarized in Fig.~\ref{feynmanfig1}. 
Diagrammatically, $\bar{P}$ and $\bar{N}$ are the sum of all Feynman
diagrams with one outgoing $P$ and $N$ line respectively.  Clearly, there
are an infinite number of diagrams contributing to $\bar{P}$ and
$\bar{N}$. These diagrams can be grouped together according to the number
of loops that they contain, thus giving rise to the loop expansion. Let
$\epsilon = 2-d $. The contribution from each diagram is a function of the
dimensionless terms $\lambda N_0 t$ and  $g(t) = \lambda t^{\epsilon/2}$
and a term that gives the correct physical dimension [$(\lambda t)^{-1}$ for
$\bar{N}$ and $(\lambda t)^{-2}$ for $\bar{P}$]. A simple combinatorial
argument shows that the contribution from a diagram with $n$ loops is
proportional to $g(t)^n$ \cite{Lee}. When $\epsilon <0$, the main
contribution to $\bar{P}$ and $\bar{N}$ comes from properly renormalized
tree level diagrams (diagrams without loops)  \cite{footnote}. When
$\epsilon>0$, the loop expansion fails since for large times $g(t)$ is no
longer a small perturbation parameter.  We therefore conclude that $2$ is
the upper critical dimension.  For $d<2$ we will use the formalism of
perturbative renormalization group to convert the loop expansion into an
$\epsilon$-expansion and calculate scaling exponents as a series in
$\epsilon$. 

\subsection{Tree level diagrams}

\begin{figure}
 \includegraphics[width=\columnwidth]{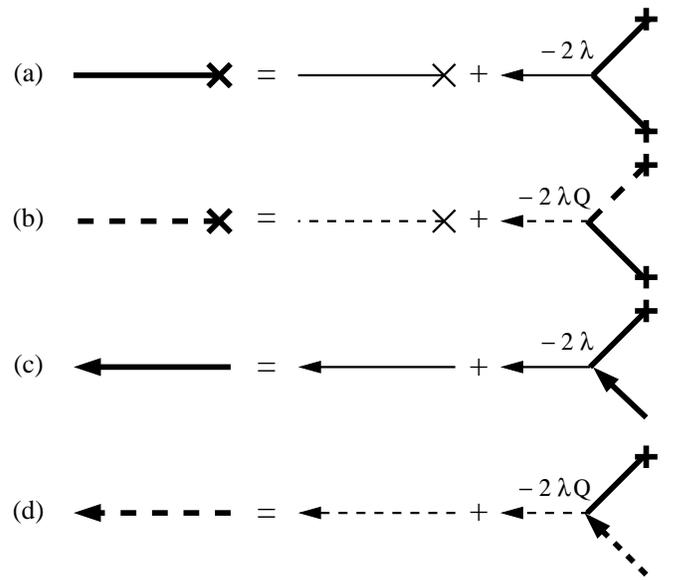}
 \caption{\label{feynmanfig2} Diagrammatic form of mean field equations
for (a) mean particle density $\bar{N}$, (b) mean density of mass $1$
particles $\bar{P}$, (c) $G_{\rm{mf}}^{\rm{NN}}$, and (d) 
$G_{\rm{mf}}^{\rm{PP}}$.}
 \end{figure}

Let $\bar{N}_{\rm{mf}}$ and $\bar{P}_{\rm{mf}}$ be mean field densities
given by the sum of contributions coming from tree diagrams with a single
outgoing $N$-line and $P$-line respectively.  We denote
$\bar{N}_{\rm{mf}}$ and $\bar{P}_{\rm{mf}}$ by thick solid lines and thick
dashed lines respectively. The integral equations satisfied by
$\bar{N}_{\rm{mf}}$ and $\bar{P}_{\rm{mf}}$ are presented in diagrammatic
form in Fig.~\ref{feynmanfig2}(a)  and \ref{feynmanfig2}(b). After
differentiating with respect to time, they can be written in analytic form
as
 \bea
 \partial_{t} \bar{P}(t)&=& -2 Q \lambda \bar{P}(t) \bar{N}(t),
\label{Nmfeq}\\
 \partial_{t} \bar{N}(t)&=& -\lambda \bar{N}^2(t) \label{Pmfeq},
 \eea
 in which one can easily recognize the Smoluchowski rate equations of the
model, obtained from Eqs.~(\ref{ssen}) and (\ref{ssep}) by neglecting the
noise terms in the right hand side. 

Equations~(\ref{Nmfeq}) and (\ref{Pmfeq}) are easily solved yielding
 \bea
 \bar{N}_{\rm{mf}}(t) &=& \frac{N_{0}}{1+\lambda N_{0}t}, \label{Nmf} \\
 \bar{P}_{\rm{mf}}(t) &=& \frac{P_{0}}{(1+\lambda N_{0}t)^{2Q}}.
\label{Pmf}
 \eea 
 From Eq.~(\ref{Pmf}), we obtain
 \be
 \theta_{\rm{mf}}=2 Q. \label{mf}
 \ee
 where $\theta_{\rm{mf}}$ is the mean field answer for $\theta$. 

In calculating loop corrections to any given order, we are faced with the
problem of summing over infinitely many diagrams containing a given number
of loops. This problem can be simplified by introducing mean field
propagators which are sums of all tree diagrams with one incoming line and
one outgoing line. Expressed in terms of these mean field propagators,
there are only finitely many diagrams with a fixed number of loops. 

Let $G_{\rm{mf}}^{\rm{NN}}$ and $G_{\rm{mf}}^{\rm{PP}}$ be mean field
propagators.  The integral equations satisfied by them are presented in
diagrammatic form in Figs.~\ref{feynmanfig2}(c)  and \ref{feynmanfig2}(d).
The solutions to these equations are
 \bea
 G_{\rm{mf}}^{\rm{NN}}(2|1) &= &\left(
\frac{\bar{N}_{\rm{mf}}(t_{2})}{\bar{N}_{\rm{mf}}(t_{1})} \right)^2
G_{0}(2|1), \label{GNN} \\
 G_{\rm{mf}}^{\rm{PP}}(2|1) & =& \left(
\frac{\bar{N}_{\rm{mf}}(t_{2})}{\bar{N}_{\rm{mf}}(t_{1})} \right)^{2Q}
G_{0}(2|1), \label{GPP}
 \eea 
 where $1=(\xv _{1},t_{1})$, $2=(\xv _{2}, t_{2})$ and $G_{0}$ is the
Green's function of the linear diffusion equation. 

\subsection{\label{sec4b}One loop diagrams}

\begin{figure}
 \includegraphics[width=\columnwidth]{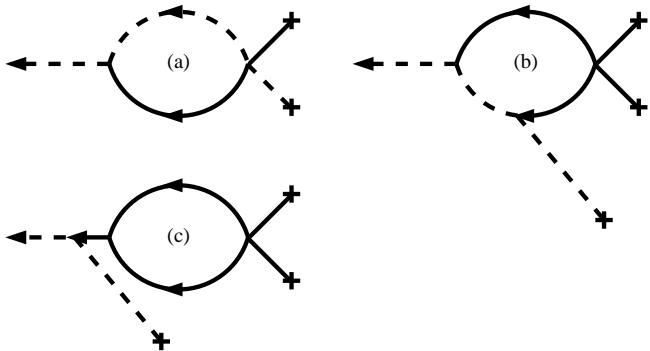}
 \caption{\label{feynmanfig3} One loop corrections to the mean field
result for $\bar{P}$.}
 \end{figure}

Using the mean field propagators and densities, it is easy to classify all
one loop diagrams contributing to $\bar{P}(t)$. These are shown in
Fig.~\ref{feynmanfig3}. The computation of the corresponding Feynman
integrals is straightforward.  The contributions from one-loop diagrams in
the limit $N_{0} \rightarrow \infty$ are
 \bea
 (a) &=& \frac{ 32 Q \lambda P_0 t^{\epsilon/2}}{(8 \pi)^{d/2} (N_0
\lambda t)^{2 Q} \epsilon^2 (\epsilon+2)}, \\
 (b) &=& \frac{ -64 Q^2 \lambda P_0 t^{\epsilon/2}}{(8 \pi)^{d/2} (N_0
\lambda t)^{2 Q} \epsilon (\epsilon+2) (\epsilon+4)}, \\
 (c) &=& \frac{ -256 Q \lambda P_0 t^{\epsilon/2}}{(8 \pi)^{d/2} (N_0
\lambda t)^{2 Q} \epsilon^2 (\epsilon+2)^2 (\epsilon+4)}, 
 \eea
 where $(a)$, $(b)$ and $(c)$ refer to the contributions from diagrams in
Figs.~\ref{feynmanfig3}(a), \ref{feynmanfig3}(b) and \ref{feynmanfig3}(c)
respectively.  Adding these one-loop contributions to the mean field
answer Eq.~(\ref{Pmf}), we obtain in the limit $N_0 \rightarrow
\infty$,
 \bea
 \lefteqn{ \bar{P}(t)= \frac{A}{t^{2 Q}} + \frac{32 Q \lambda A} {(8
\pi)^{d/2} \epsilon t^{2 Q - \epsilon/2}} \left[ \frac{\epsilon+6-2 Q
(\epsilon+2)} {(\epsilon+2)^2 (\epsilon+4)} \right]} \qquad \qquad
\nonumber \\
 && \mbox{} + \mbox{$2$- and higher loop corrections,} \label{oneloop}
 \eea
 where $A=P_0/(N_0 \lambda)^{2 Q}$. 

\subsection{\label{sec4c}Renormalization Group Analysis of the Model.}

The large time asymptotic behavior of $\bar{P}(t)$ can be obtained by
solving the Callan-Symanzik equation with initial conditions given by
Eq.~(\ref{oneloop}) (see Ref.~\cite{cardy} for a review). The coefficients
of Callan-Symanzik equation are determined by the law of renormalization
of all the relevant couplings of the theory Eqs.~(\ref{ssen}) and
(\ref{ssep}). Power counting analogous to that carried out in \cite{KRZ1},
shows that there are only two relevant couplings of the theory in $d<2$:
the reaction rate $\lambda$ and the initial mass distribution $P_0$. We
will derive the one-loop renormalization law of the initial mass
distribution by requiring that Eq.~(\ref{oneloop}) is non-singular in the
limit $\epsilon \rightarrow 0$ if expressed in terms of renormalized
relevant couplings. 

Let $t_0$ be a reference time and $g_0= \lambda t_0^{\epsilon/2}$ be the
dimensionless reaction rate. We choose $t_{0}$ in such a way that $g_{0}
\ll 1$. The mechanism of renormalization of the reaction rate in the
theory is identical to that of the reaction $A+A\rightarrow A$. 
Physically, the renormalization of reaction rate is explained by the
recurrent property of random walks. The probability of a reaction between
particles at time $t$ is proportional to the ''bare'' reaction rate,
multiplied by the probability that the reaction has not occurred before
time $t$. In $d \leq 2$, the latter probability explicitly depends on time
$t$. The law of renormalization of the reaction rate has been worked out
in Ref.~\cite{Peliti}. If $g_{R}$ is the renormalized reaction rate, then
it is related to $g_0$ by the relation
 \be
 g_{R} = \frac{g_0}{1+g_0/ g^*},
 \label{gR}
 \ee
 where $g^* = 2 \pi \epsilon + O(\epsilon^2)$ is the nontrivial fixed
point of the renormalization group flow in the space of effective coupling
constants. The mass distribution $\bar{P}(t_0)$ can now be expressed in
terms of the renormalized reaction rate $g_{R}$ to be
 \be
 \bar{P}(t_0) = \frac{P_0}{(N_0 g_{R} t_0^{d/2})^{2 Q}} \left[ 1 -
\frac{g_R}{g^*} Q (2 Q-1) + O(g_R^2) \right]. \label{stage1}
 \ee

The order $g_R$ term in Eq.~(\ref{stage1}) is singular at $\epsilon=0$. 
To cancel this divergence, we have to introduce a renormalized initial
mass distribution $P_R$:
 \be
 P_R = Z(g_R,t_0,\epsilon) P_0,
 \ee
 where $Z(g_R,t_0,\epsilon)$ is chosen such that
 \be
 \bar{P}(t,g_R,P_R,t_0) = Z(g_R, t_0, \epsilon) \bar{P}(t,\lambda, P_0,
\epsilon), \label{Zdef}
 \ee
 is non-singular at $\epsilon=0$. Substituting Eq.~(\ref{Zdef}) into
Eq.~(\ref{stage1}), we obtain
 \be
 Z = 1+\frac{g_R}{g^*} Q (2 Q -1) + O(g_R^2). 
 \ee

The Callan-Symanzik equation is obtained by noting that
$\bar{P}(t,\lambda,P_0, \epsilon)$ does not depend on the reference time
$t_0$. Therefore,
 \be
 t_0 \frac{\partial}{\partial t_0} \left[Z^{-1} \bar{P}(P_R) \right] = 0. 
\label{CS1}
 \ee
 It follows from dimensional analysis that the most general form
 of the average mass distribution is
 \be
 \bar{P}(t) = \frac{1}{t_0^{d Q}} \frac{P_R}{N_0^{2 Q}}
\Phi\left(\frac{t}{t_0}, g_R \right), \label{scaling}
 \ee
 where $\Phi$ is a dimensionless function. Using the scaling function
Eq.~(\ref{scaling}) in Eq.~(\ref{CS1}), we obtain the Callan-Symanzik
equation for $\bar{P}(t)$: 
 \be
 \left( t \frac{\partial}{\partial t} + \frac{\beta(g_R)}{2}
\frac{\partial}{\partial g_R} + d Q - \frac{\gamma(g_R)}{2} \right) 
\bar{P}(t,g_R,P_R,t_0)=0, \label{CS2}
 \ee
 where
 \bea
 \beta(g_R) &=& -2 t_0 \frac{\partial g_R}{\partial t_0} = \frac{g_R (g_R
- g^*) \epsilon}{g^*}, \label{beta} \\
 \gamma(g_R) &=& \frac{-2 t_0}{Z} \frac{\partial Z}{\partial t_0} =
\frac{-2 Q (2Q - 1)}{4 \pi} g_R + O(g_R^2), \label{gamma}
 \eea
 are the beta and gamma functions of the theory. 

At large times, the solutions of Eq.~(\ref{CS2}) are governed by the
nontrivial fixed point $g^*$ of the beta function. It then follows from
the Callan-Symanzik Eq.~(\ref{CS2}) that $\bar{P}(t) \sim t^{-\theta}$
where
 \be
 \theta = Q d - \frac{\gamma(g^*)}{2}. \label{dstar}
 \ee
 Thus we obtain from Eqs.~(\ref{gamma})  and (\ref{dstar}) that
 \be
 \theta = d Q + Q (Q - \frac{1}{2}) \epsilon + O(\epsilon^2), ~\epsilon >
0. 
 \label{thetaoneloop}
 \ee

The knowledge of $\theta$ to the first order in $\epsilon$ combined with
Eqs.~(\ref{beta1}) and (\ref{gamma1}) allows one to calculate the
exponents $\delta$ and $\zeta$ with the same precision: 
 \bea
 \delta &=& d (1-Q) + (Q-\frac{1}{2}) (Q-1) \epsilon + O(\epsilon^2),
\label{delta1loop} \\
 \zeta &=& (2 Q-1) \ep + O(\ep^2). \label{gamma1loop}
 \eea
 The exponent $\zeta$ is proportional to the sum of the anomalous
dimensions of $P$ and $\rho$. As a result, the mass dependence of $\pb$
can be captured neither by mean field theory nor by Smoluchowski
approximation (see \cite{KRZ1} for a more detailed discussion of this
point). 

The results of this subsection can be summarized as follows. The mean mass
distribution $\bar{P}(m,t)$ varies as
 \be
 \pb \sim \frac{(m^{1/d})^{(2Q-1) \epsilon + O(\epsilon^2)}}{t^{d Q + Q (Q
-1/2) \epsilon + O(\epsilon^2)}}, \label{answer}
 \ee
 for $m \ll M(t)$, where $M(t)$ is the typical mass at time $t$, or 
equivalently the typical number of coagulations
undergone by all
ancestors of survived particles. $\bar{P}(m,t)$ decays algebraically with
time with an exponent independent of $m$. The coefficient multiplying this 
time dependent term
does however grow algebraically with $m$. 

\subsection{\label{sec4d}Two dimensions.}

The upper critical dimension of our model is $2$. The non-trivial fixed
point of the $\beta$-function Eq.~(\ref{beta}) vanishes at $d \rightarrow
2$. We therefore expect the mean field answers Eqs.~(\ref{Nmf}) and
(\ref{Pmf}) to give the correct large time-small mass of average densities in
two dimension, modulo logarithmic corrections. In this subsection
we calculate these corrections. 

When $Q=1$ it was shown that in two dimensions, $\bar{P}(m,t)\sim \ln(t)
\ln(m) t^{-2}$ for $t \rightarrow \infty, m \ll M(t)$ \cite{KRZ1}. To
calculate these corrections for arbitrary $Q$ we need to solve
the Callan-Symanzik equation (\ref{CS2}) with coefficients calculated at
$d=2$. In two dimensions,
 \bea
 \beta(g)|_{d=2} &=& \frac{g^2}{2 \pi}, \\
 \gamma(g)|_{d=2} &=& \frac{-2 Q(2 Q-1)g}{4 \pi} + O(g^2). 
 \eea
 Then Eq.~(\ref{CS2}) reduces to
 \be
 \left( t \frac{\partial}{\partial t} + \frac{g_R^2}{4 \pi}
\frac{\partial}{\partial g_R} + 2 Q + \frac{Q(2 Q-1)g_R}{4 \pi} \right) 
\bar{P}(t,g_R,t_0)=0, \label{CS2d}
 \ee
 which has to be solved with the initial condition
 \be
 \bar{P}(t_0) = \frac{\mbox{const}}{(g_R t_0)^{2Q}},
 \ee
 provided by the mean field theory. The solution to Eq.(\ref{CS2d}) with
this initial condition is
 \be
 \bar{P}(t) = {\mbox{const}} \times \frac{\left(\ln(t/t_0)\right)^{Q(3-2
Q)}} {g_R^{Q (2 Q-1)}t^{2 Q}} \left(1+ O (\frac{1}{\ln(t/t_0)}) \right).
\label{pb2d}
 \ee 
 When $Q=1$, we recover the result of Ref.~\cite{KRZ1}. When $Q=0$, $P(t)$
ceases to depend on time, as it should. When $Q=1/2$, $\bar{P}\sim
\ln(t)/t$, which coincides with the decay law of the concentration of
particles in $A+A\rightarrow \emptyset$ reaction \cite{Peliti}.

The dimensional arguments that led to Eq.~(\ref{gamma1})  cannot capture
the mass dependent
logarithmic corrections that are present in $2$-dimensions. Hence, we
need to generalize these dimensional arguments. This is provided by the
Callan-Symanzik equation obeyed by $\bar{P}(m,t)$ when considered as a
function of both $m$ and $t$. 

The full distribution $\pb$ cannot depend on the choice of reference time
$t_{0}$, which we introduced to regularize the perturbative expansion of
$\bar{P}(t)$. Therefore,
 \be
 t_{0}\frac{\partial \pb}{\partial t_{0}}=0. \label{gcs1}
 \ee
 From dimensional analysis, it follows that
 \be
 \pb=\frac{\bar{N}(t_{0})^2}{\bar{\rho}(t_{0})} F\bigg(
\frac{m\bar{N}(t_{0})}{\bar{\rho}(t_0)},\frac{t}{t_{0}},
g_{R}\bigg).\label{pcomplete}
 \ee
 The form in Eq.~(\ref{pcomplete}) is different from the scaling form used
in Eq.~(\ref{scaling}) because $P_R$ has to be now expressed in terms of
$m$.  Substituting Eq.~(\ref{pcomplete}) into Eq.~(\ref{gcs1}) we obtain
 \be 
 \left[(d_{\rho}\!\!-\!d_{N})m\frac{\partial}{\partial m}-
\!t\frac{\partial}{\partial t}-\frac{\beta(g_R)}{2}
\frac{\partial}{\partial g_R} +(d_{\rho}\!\!- \!2d_{N})\right]F=0,
\label{gcs2}
 \ee
 where $d_{\rho}=2(1-Q)- (1-Q)(1+2Q) g_R/(4 \pi)+O(g_R^2)$ and
$d_{N}=1-g_R/(4 \pi)+O(g_R^2)$ are scaling dimensions of fields $N$ and
$\rho$, which can be obtained from the corresponding loop expansions. 

We look for solutions of Eq.~(\ref{gcs2}) of the form
 \be
 F=F_1\left( \frac{t}{t_{0}},g_R, t_{0} \right)  F_2\left( \frac{m
\bar{N}(t_{0})}{\bar{\rho}(t_{0})}, g_R, t_{0} \right). 
 \label{product}
 \ee
 The time dependent function $F_1$ obeys the Callan-Symanzik
equation~(\ref{CS2}). Using this fact and substituting Eq.~(\ref{product})
into Eq.~(\ref{gcs2}), we obtain
 \be 
 \left[ m\frac{\partial }{\partial m} +\frac{1+O(g_R)}{2(2Q-1)}
\beta(g_{R})\frac{\partial }{\partial g_{R}}-\Gamma (g_{R})\right] F_2 =
0, \label{gcs3}
 \ee
 where $\Gamma (g_R)=(2Q-1)g_{R}/(2 \pi d)+O(g_{R}^2)$. As expected, when
$d<2$, $Q>\frac{1}{2}$ and $t\rightarrow \infty$ the solution of
Eq.~(\ref{gcs3}) is given by Eq.~(\ref{answer}). Let $d=2$. Solving
Eq.~(\ref{gcs3}) for $Q\geq \frac{1}{2}$ with the initial condition
$F_2(m_{0})=\rm{const}$, provided by mean field theory, we find that
 \be 
 F_2(m)\sim \left(\ln(\frac{m}{m_{0}})\right)^{(2Q-1)^2}\bigg( 1+O\big(
1/\ln(m/m_{0}) \big) \bigg), \label{B}
 \ee 
 where $m_{0}=\frac{\bar{N} (t_{0})}{\bar{\rho} (t_{0})}$ is a reference
point in mass space.

Combining Eqs.~(\ref{pb2d}) and (\ref{B}), we conclude that in $d=2$
 \be
 \pb \sim \frac{\ln(t)^{Q(3-2Q)}\ln(m)^{(2Q-1)^2}}{t^{2Q}},
 \ee
 for $m_{0} \ll m \ll M(t)$ and $t\rightarrow \infty$. For $Q=1$ we
recover the answer for the average mass distribution in the $A_i+A_j
\rightarrow A_{i+j}$ model obtained in Ref.~\cite{KRZ1}.  This result has
also been verified numerically \cite{KRZ1}.  When $Q=1/2$, $\pb$ no longer
depends on mass, as expected. 

\subsection{\label{sec4e}Comparison with results from Monte Carlo
simulations}

In this subsection, we compare the results obtained for the exponents
$\theta$ and $\delta$ as an $\epsilon$-expansion with results from Monte
Carlo simulations in $1$-dimension. Let $\theta_1$ denote the value of
$\theta$ obtained by truncating the $\epsilon$-expansion at order
$\epsilon$ and setting $\ep=1$. Then,
 \be
 \theta_1= \frac{Q}{2}+Q^2. 
 \label{theta1}
 \ee
 In Table~\ref{table1}, we compare this analytic expression with results
from numerical simulations (see columns $2$ and $3$). There is good
agreement.
 \begin{table}
 \caption{\label{table1} The numerically obtained values of $\theta$ for
different values of $q$ are compared with $\theta_1$ (Eq.~(\ref{theta1}))
and $\theta_2$ (Eq.~(\ref{pecm})).  For $q<2$, the numerical values are
obtained by measuring the decay of mean density and then using
Eq.~(\ref{beta1}) . For $q>2$, the numerical results are from
Refs.~\cite{MC} and \cite{DB2}.}
 \begin{ruledtabular}
 \begin{tabular}{rlcc} $q$ & Numerical & $\theta_1$ & $\theta_2$ \\ \hline
 1.11  & $0.08 \pm 0.01$  & 0.06  &0.08 \\
 1.25  & $0.18 \pm 0.01$  & 0.14  &0.17 \\
 1.50  & $0.32 \pm 0.01$  & 0.28  &0.30 \\
 1.77  & $0.41 \pm 0.01$  & 0.41  &0.42 \\
 2.00  & $0.50$  & 0.50  &0.50 \\
 3.00  & $0.73 \pm 0.01$  & 0.78  &0.75 \\
 4.00  & $0.87 \pm 0.01$  & 0.94  &0.91 \\
 5.00  & $0.96 \pm 0.01$  & 1.04  &1.01 \\
 6.00  & $1.04 \pm 0.01$  & 1.11  &1.08 \\
 8.00  & $1.12 \pm 0.01$  & 1.20  &1.17 \\
 16.00 & $1.28 \pm 0.01$  & 1.35  &1.33 \\
 25.00 & $1.35 \pm 0.01$  & 1.40  &1.39 \\
 32.00 & $1.38 \pm 0.01$   & 1.42  &1.41 \\
 50.00 & $1.42 \pm 0.01$  & 1.45  &1.44 \\
 $\infty$ & $1.50$  & 1.50  &1.50 \\
 \end{tabular}
 \end{ruledtabular}
 \end{table}

To go beyond the expression in Eq.~(\ref{theta1}) and to make an estimate
of the error arising by neglecting terms of order $\ep^2$ and higher, we
proceed as follows. Let the corrections from order $\ep^2$ and higher
orders be denoted by $R(\ep,Q)$, such that
 \be
 \theta = d Q + Q (Q - \frac{1}{2})\epsilon+ R(\ep,Q). 
 \ee
 $R(\ep, Q)$ must vanish at $Q=0$ and $Q=1/2$ (see Eq.~(\ref{exact})). 
Moreover, $R(1, 1)=0$, since $\theta=3/2$ when $Q=1$ and $\ep=1$
\cite{spouge}. Therefore,
 \be
 R(\ep,Q)= \ep^2 Q (Q-\frac{1}{2})\left[\!(1-Q) h_1(\ep,Q)+(\ep -1)
h_2(\ep,Q)\right],
 \ee
 where $h_1$ and $h_2$ are unknown functions.  Setting $\ep=1$, we obtain
 \be
 \theta = \frac{Q}{2} + Q^2 + Q (Q-\frac{1}{2})(Q-1) h_1(1,Q), \label{pe}
 \ee
 
The value of function $h_1(1,Q)$ at $Q=0$ can be determined. It was
shown in Ref.~\cite{monthus} that
 \bea
 \theta =\frac{3\sqrt{3}}{2\pi} Q + O(Q^2). 
 \label{cm}
 \eea
 Therefore, $h_1(1,0)=3\sqrt{3}/\pi-1$.  A two loop calculation carried out
in Sec.~\ref{sec5} shows that the function $h_1(1,Q)$ is slowly varying in
the interval $Q\in [0,1]$. Therefore, we replace the function $h_1(1,Q)$ by
its value at $Q=0$ and denote the resulting expression as $\theta_2$. 
Thus, we obtain
 \be
 \theta_2= \frac{Q}{2}+ Q^2 +Q(Q-\frac{1}{2})(Q-1) (\frac{3
 \sqrt{3}}{\pi}-1). 
 \label{pecm}
 \ee
 In Table~\ref{table1}, we compare $\theta_2$ with results from Monte
Carlo simulations in $1$-dimension. The error decreases as compared to
$\theta_1$. 

The error due to dropping terms of order $\ep^2$ and higher can be
estimated. The function $h_1(1,Q)$ in Eq.~(\ref{pe}) is or order $1$.  The
function $|Q(Q-1/2)(Q-1)|$ takes on a maximum value of $0.05\ldots$ in the
interval $Q \in [0.5,1]$. Hence the absolute error is of order $0.05$,
which is in agreement with the results presented in Table~\ref{table1}. 

We do a similar analysis for $\delta$. Let $\delta_1$ be the value of
$\delta$ obtained by truncating the series Eq.~(\ref{delta1loop}) and then
putting $\ep=1$. Then
 \be
 \delta_1 = \frac{3}{2} - \frac{5 Q}{2} +Q^2. 
 \label{delta1}
 \ee
 To obtain $\delta_2$, we substitute $Q\rightarrow (1-Q)$ in
Eq.~(\ref{pecm}) to obtain
 \be
 \delta_2 = \frac{3}{2} - \frac{5 Q}{2} +Q^2
 -Q(Q-\frac{1}{2})(Q-1) (\frac{3 \sqrt{3}}{\pi}-1). 
 \label{delta2}
 \ee
 In Table~\ref{table2}, we compare the results $\delta_1$ and $\delta_2$
with results from numerical simulations. Very good agreement is seen. 
 \begin{table}
 \caption{\label{table2} The numerically obtained values of $\delta$ for
different values of $q$ are compared with $\delta_1$ (Eq.~(\ref{delta1})) 
and $\delta_2$ (Eq.~(\ref{delta2})). }
 \begin{ruledtabular}
 \begin{tabular}{rlcc}
 $q$ & Numerical & $\delta_1$ & $\delta_2$ \\ \hline
 2  & $0.50$  & 0.50  &0.50 \\
 3  & $0.31 \pm 0.01$  & 0.28  &0.30 \\
 4  & $0.22 \pm 0.01$  & 0.19  &0.22 \\
 5  & $0.18 \pm 0.01$  & 0.14  &0.17 \\
 8  & $0.11 \pm 0.01$  & 0.08  &0.10 \\
 16 & $0.05 \pm 0.01$  & 0.04  &0.05 \\
 $\infty$ & $0.00$  & 0.00  &0.00 \\
 \end{tabular}
 \end{ruledtabular}
 \end{table}

\section{\label{sec5}The analysis of two- and higher loop corrections.}

\subsection{\label{sec5a}General structure of the loop expansion.}

In this subsection, we examine the contributions from diagrams with $2$-
and more loops. It will be shown that the coefficient of $\epsilon^n$ in
the $\epsilon$-expansion of $\theta$ is a polynomial of degree $2 n$ in
$Q$. It is easier to derive the result, not by using the formalism of
renormalization group, but by identifying the principal set of diagrams
contributing to the large time limit of $\bar{P}(t)$ and deriving a simple
integral equation satisfied by the sum of these diagrams. 

\begin{figure}
 \includegraphics[width=\columnwidth]{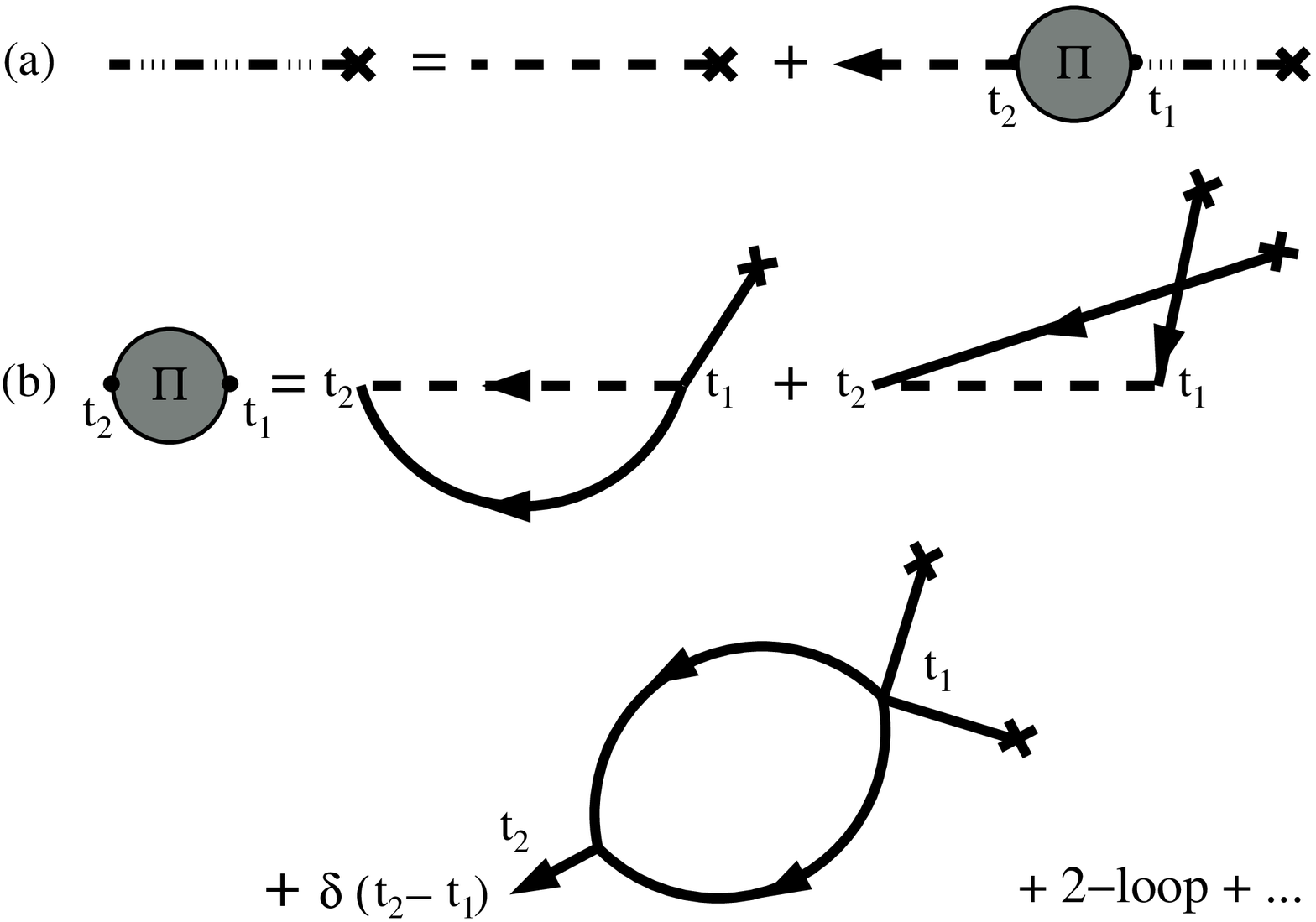}
 \caption{\label{feynmanfig4} (a) Schwinger-Dyson equation for
$\bar{P}(t)$. (b) Perturbative expansion of the polarization operator
$\Pi$.}
 \end{figure}

The polarization operator $\Pi(t_2,t_1)$ is defined as the sum of all
one-particle irreducible diagrams with one outgoing and one incoming
$P$-line, with the external propagator lines stripped off.  Using the
polarization operator, we can write down the Schwinger-Dyson equation
obeyed by $\bar{P}(t)$.  Let $\bar{P}(t)$ and $\Pi(t_2,t_1)$ be denoted by
a thick dashed-dotted line and by a grey circle respectively. Then,
$\bar{P}(t)$ satisfies the equation shown diagrammatically in
Fig.~\ref{feynmanfig4}(a). In equation form, it is
 \be
 \bar{P}(t) = \bar{P}_{\rm{mf}}(t) + \frac{1}{t^{2 Q}} \int_0^{t} dt_2~
t_2^{2 Q} \int_0^{t_2} dt_1 \Pi (t_2,t_1) \bar{P}(t_1). \label{dyson}
 \ee
 In $2$-dimensions, $\bar{P}(t) \sim t^{-2 Q}$.  Let $\eta(t) = t^{2 Q}
\bar{P}(t)$.  In terms of $\eta$, Eq.~(\ref{dyson}) reduces to
 \be
 \eta(t) = \eta_0 + \int_0^t dt_2 \int_0^{t_2} dt_1 \left[t_2^{2 Q}
\Pi(t_2,t_1) t_1^{-2 Q} \right] \eta(t_1),
 \ee
 where $\eta_0$ is a constant independent of $t$. Differentiating with
respect to $t$, we obtain
 \be
 \frac{d \eta(t)}{d t} = \int_0^{t} dt_1 \left[t^{2 Q} \Pi(t,t_1) t_1^{-2
Q} \right] \eta(t_1). \label{sd}
 \ee

The expansion of $\Pi(t,t_1)$ in terms of Feynman diagrams is shown in
Fig.~\ref{feynmanfig4}(b). From the Feynman diagrams in
Fig.~\ref{feynmanfig1}, it follows that the number of dotted lines in any
given diagram is conserved as one moves from the right to the left. Any
diagram contributing to the polarization operator has a single dotted line
threading through it from right to left. Also, the only vertex that
contributes a factor $Q$ is the $\rm{PPN}$ vertex. Therefore, the
$Q$-dependent part of a given diagram has the form $Q^{\# \mbox{of PPN
vertices}} \prod_{i}(t_i/t_{i+1})^{2Q}$, where the product is over all
vertices involving the $P$-line. For a $k$-loop diagram, we can have
utmost $2 k$ vertices of the type $\rm{PPN}$.  Also, it was shown in
Ref.~\cite{KRZ1} that $n$-loop diagrams contribute at the order
$\epsilon^n$ only. Hence, after coupling constant renormalization, one
finds that the expression in square brackets of Eq.~(\ref{sd}) is of the
form $t^{-2} \sum_{n=1}^{\infty}\ep^{n}P_{2n}(Q)$, where $P_{2n}(Q)$ is a
polynomial of degree $2n$ in $Q$, and where the factor $t^{-2}$ has been
pulled out to give the right dimension. 

We can now solve Eq.~(\ref{sd})  perturbatively, order by order in
$\epsilon$. Simple dimension counting shows that $t^{2 Q} \Pi(t,t_1)
t_1^{-2 Q} = t^{-2} F(t_1/t)$, where $F(\tau)$ is a dimensionless
function. The previous argument shows that $F(\tau) =
\sum_{n=1}^{\infty}\ep^{n}P_{2n}(Q)$. Assume a power law solution for
$\eta$, i.e., $\eta= c t^{-\theta_p}$ where $\theta_p= \sum_{n=1} a_n
\epsilon^n$ and $c$ is a constant.  Then Eq.~(\ref{sd}) simplifies to
 \be
 \theta_p = -\int_0^1 d \tau F(\tau) \tau^{-\theta_p}. 
 \ee
 Expanding $F(\tau)$ and $\theta_p$ as series in $\epsilon$, we obtain
 \be
 \sum_{n=1}^{\infty} a_n \epsilon^n = -\int_0^1 d \tau
\sum_{n_1=1}^{\infty} P_{2 n_1} \epsilon^{n_1} \sum_{n_2=0}^{\infty}
\frac{\left[-\theta_p \ln(\tau)\right]^{n_2}}{n_2!}. 
 \label{perturbative}
 \ee
 Solving Eq.~(\ref{perturbative}) order by order for $a_n$, it is easy to
verify that the coefficient of $\epsilon^n$ is a polynomial of degree $2
n$ in $Q$, i.e.,
 \be
 \theta=\sum_{n=0}^{\infty}\ep^{n}\left(\sum_{p=0}^{2 n}
C_{n,p}Q^{p}\right), \label{struc}
 \ee
 where $C_{n,p}$'s are some unknown constants. 

Given Eq.~(\ref{struc}), it is easy to rederive the one loop correction to
$\theta$ (Eq.~(\ref{thetaoneloop})) obtained by the renormalization group
formalism. The three unknowns in the coefficient of $\epsilon$ in
Eq.~(\ref{struc}) are obtained from the exact results in Eq.~(\ref{exact})
giving $C_{1,0}=0$, $C_{1,1}= -3/2$, and $C_{1,2} = 1$. 

\subsection{\label{sec5b}Two-loop formula for $\zeta(Q)$.}

As the mean field answer for the exponent $\zeta$ is $0$, it is desirable
to evaluate order $\ep^2$ correction to Eq.~(\ref{gamma1loop}). This
requires the knowledge of order $\ep^2$ term in $\theta$. 

From Sec.~\ref{sec5a}, we know that the term in $\epsilon^2$ is a
polynomial of degree $4$ in $Q$, i.e.,
 \be
 \theta=d Q+ Q(Q- \frac{1}{2}) \epsilon + \left( \sum_{k=0}^{4} C_{2,k}
Q^k \right)\ep^2+O(\ep^3). 
 \ee
 Out of the $5$ unknown $C_{2,k}$'s, two are fixed by the conditions that
$\theta=0$ for $Q=0$ and $\theta=d/2$ for $Q=1/2$ (see Eq.~(\ref{exact}).
For $Q=1$, it is known that $\theta=d+\ep/2+ O(\ep^2)$ \cite{KRZ1}. We
assume that the order $\ep^2$ term is absent when $Q=1$ (see Appendix for
a heuristic validation of this assumption). This fixes the third constant
and we are left with
 \be
 \theta=d Q+ Q(Q- \frac{1}{2}) \epsilon +Q(Q- \frac{1}{2})(Q-1)
(AQ+B)\ep^2+O(\ep^3),
 \label{pe2}
 \ee
 where $A$ and $B$ are constants. 
 
\begin{figure}
 \includegraphics[width=\columnwidth]{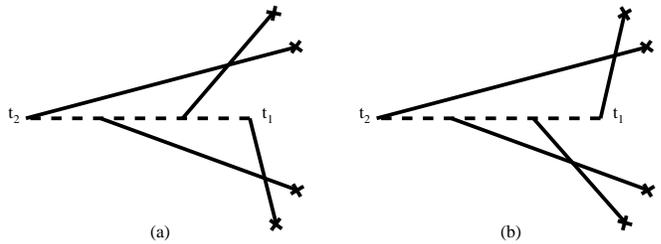}
 \caption{\label{feynmanfig5} The two-loop diagrams contributing to the
constant $A$ in Eq.~(\ref{pe2}).}
 \end{figure}

The constant $A$ is not difficult to calculate. The contribution to
$\theta$ of order $\ep^2Q^4$ comes from the square of 1-loop polarization
operator and from two-loop diagrams with four $PPN$-vertices. There are
only two such diagrams, which are shown in Fig.~\ref{feynmanfig5}. The
numerical computation of corresponding Feynman integrals gives
 \be
 A=(\frac{\pi^2}{3}-3)-0.296 \ldots \approx -0.006, \label{A}
 \ee
 where the first term on the right hand side comes from the order-$\ep^2$
term in the one-loop polarization operator. 

The calculation of constant $B$ seems an almost impossible task, as there
are over twenty two-loop diagrams contributing to it. However, the terms
proportional to B drop out of the the two loop expression for $\zeta$.
Substituting Eq.~(\ref{pe2}) into Eq.~(\ref{gamma1}) one finds that
 \be
 \zeta=(2 Q-1) \epsilon + (2 Q-1) (Q-1) (\frac{1}{2}+ A Q)
\epsilon^2+O(\ep^3). 
 \label{gamma2lp}
 \ee
 In Table~\ref{table3}, we compare the one-loop expression for $\zeta$
(Eq.~(\ref{gamma1loop})) and two-loop expression for $\zeta$
(Eq.~(\ref{gamma2lp})) with numerical results in $1$-dimension. 
 \begin{table}
 \caption{\label{table3} Comparison of one-loop (Eq.~(\ref{gamma1loop})) 
and two-loop (Eq.~(\ref{gamma2lp})) results for $\zeta$ with numerical
simulations in $1$-dimension.}
 \begin{ruledtabular}
 \begin{tabular}{rlcc}
 $q$ & Numerical & $1$-loop & $2$-loop \\
\hline
 2  & 0.00  & 0.00  &0.00 \\
 3  & $0.21 \pm 0.11$  & 0.33  &0.28 \\
 4  & $0.32 \pm 0.08$  & 0.50  &0.44 \\
 5  & $0.44 \pm 0.08$  & 0.60  &0.54 \\
 8  & $0.59 \pm 0.07$  & 0.75  &0.70 \\
 16 & $0.73 \pm 0.06$  & 0.88  &0.85 \\
 $\infty$ & 1.00  & 1.00  &1.00 \\
 \end{tabular}
 \end{ruledtabular}
 \end{table}

\section{\label{sec6}Summary and conclusions}

In summary, we develop a systematic method to calculate the persistence
exponent $\theta$ for a system of coagulating and annihilating random
walkers, in arbitrary dimensions.  In $1$-dimension, this corresponds to
persistence probabilities of domain walls in the Potts
model evolving via zero temperature Glauber dynamics. We establish an
exponent relation by which the number of unknown exponents in the problem
is reduced from two to one. The unknown persistence exponent $\theta$
is determined
perturbatively using the formalism of renormalization group. 

The persistence problem studied in this paper can be considered as a
special case of a more general problem of the survival probability of a
test particle with diffusion constant $\kappa$ times the diffusion
constant of the other particles. In this case it is known that the
persistence exponent $\theta_{\kappa}(q)$ depends on $\kappa$
\cite{monthus,MC}.  While simple limiting cases have been studied
\cite{monthus,FG,KNR,howard} via numerics, mean-field or perturbative
techniques, a general understanding is still lacking. It would be
interesting to extend the formalism of perturbative renormalization group
to calculate $\theta_{\kappa}(q)$.

\section*{Acknowledgments}

The work at Oxford was supported by EPSRC, UK. We would like to thank
Satya Majumdar, Alan Bray and John Cardy for useful discussions.

\appendix

\section{\label{appendix1}Mass distribution in the $A_{i}+A_{j}
\rightarrow A_{i+j}$ model}

In this appendix, we present a heuristic derivation of the distribution of
small masses in the $A_{i}+A_{j} \rightarrow A_{i+j}$ model. This model
corresponds to the $Q=1$ limit of the model discussed in the paper. For
the $A_i + A_j \rightarrow A_{i+j}$ model, it is known \cite{spouge,KRZ1}
that for $m \ll t^{d/2}$,
 \be
 \bar{P} (m,t)  \sim \cases{
 \frac{m}{t^{3/2}} & in $d=1$, \cr
 \frac{m^{(\epsilon+O(\epsilon^2))/d}}{t^{d+\epsilon/2 +O(\epsilon^2)}} &
in $1\leq d < 2$, \cr
 \frac{\ln(m) \ln(t)}{t^2} & in $d = 2$, \cr
 \frac{1}{t^2} & in $d > 2$. \cr}
 \label{a1}
 \ee
 In this appendix, we give a heuristic argument as to why the terms of order
$\epsilon^2$ and higher could be absent, as a result of which 
the expansion up to order
$\epsilon$ gives the exact answer. 

Putting $\lambda_a=0$ in Eq.~(\ref{sse}), we obtain that the mass
distribution $P(m,x,t)$ evolves according to
 \be
 \left(\frac{\partial}{\partial t} - D \nabla^2 \right) P = \lambda_c P*P
- 2 \lambda_c N P + i \sqrt{2 \lambda_c} \xi P,
 \ee
 where $N= \int_0^{\infty} dm P(m)$ is the density of particles 
and $P*P= \int_0^m dm' P(m') P(m-m')$. The density obeys the equation: 
 \be
 \left(\frac{\partial}{\partial t} - D \nabla^2 \right) N = \lambda_c N^2
+ i \sqrt{2 \lambda_c} \xi N. 
 \ee Let
 \be
 F(s,t) = \int_0^\infty dm P(m,t) e^{-m s}
 \ee 
 be the Laplace transform of $P(m,t)$. Then,
 \be
 B(s,t)= N(t) - F(s,t),
 \ee
 obeys the equation
 \be
 \left(\frac{\partial}{\partial t} - D \nabla^2 \right) B = \lambda_c B^2
+ i \sqrt{2 \lambda_c} \xi B. 
 \ee

The function $B(s,t)$ obeys the same equation as the density $N$
\cite{oleg}.
However, the initial conditions at $t=0$ are different. If the initial density
of particles $N(0)=N_0$, then $B(s,0)=N_0 (1-e^{-s})$. Therefore, if the
average particle density is $\bar{N}(N_0,t)$, then
 \be
 \bar{F}(s,t) = \bar{N}(N_0,t)- \bar{N}(N_0 (1-e^{-s}),t). 
 \label{a7}
 \ee
 The function $F(s,t)$ will have the scaling form
 \be
 \bar{F}(s,t) \sim \frac{1}{t^{d/2}} g (s t^{d/2}),
 \ee
 where the scaling function $g(x) \sim x^{-\phi}$ for $x \gg 1$. On
performing the inverse Laplace transform, we obtain
 \be
 \bar{P}(m,t) \sim \frac{m^{\phi-1}}{t^{d/2 (1+\phi)}}. 
 \label{a9}
 \ee
 Thus, if $\phi$ were equal to $2/d$, the expression in Eq.~(\ref{a1}) to
is exact order $\ep$. 

To calculate $\phi$, we look at the behavior of the particle density 
$\bar{N}$. It is expected to have the scaling form
 \be
 \bar{N} = N_0 h(N_0 t^{d/2}) 
 \label{a10}
 \ee
 where the scaling function $h(x)$ behaves for large $x$ as
 \be
 h(x) \sim \frac{1}{x} (1+\frac{1}{x^\psi}), \qquad x \gg 1,
 \label{a11}
 \ee
 where $\psi$ is some exponent greater than zero. 

Substituting Eqs.~(\ref{a10}) and (\ref{a11}) into Eq.~(\ref{a7}), it is
straightforward to verify that
 \be
 \phi = \psi. 
 \ee
 In $1$- and $2$-dimensions, it is easy enough to verify that $\psi=2$ and
$\psi=1$ respectively, consistent with the exact results in
Eq.~(\ref{a1}). In other dimensions, we argue as follows. The density of
particles in the $A_i + A_j \rightarrow A_{i+j}$ model is inversely
proportional to the area swept out by a random walker in time $t$. This
area varies as $(\sqrt{t+c})^d$, where $c$ is some constant. Then, the
particle density decays as $\bar{N} \sim t^{-d/2} (1- \rm{const}/t)$, such that
 \be
 \psi=\frac{2}{d}. 
 \ee
 Substituting into Eq.~(\ref{a9}), we obtain
 \be
 \bar{P}(m,t) \sim \frac{m^{(2-d)/d}}{t^{d/2 +1}}. 
 \ee
 which is the one-loop answer in Eq.~(\ref{a1}).


\begin{thebibliography}{1}
\bibitem{footnote} The effective reaction rate entering the Smoluchowski
coagulation equation is not equal to the bare reaction rate. This is due
to strong ultraviolet divergence of Feynman integrals contributing to
renormalization of the reaction rate. As a result, the density-density
correlation function is not equal to the product of one-point functions,
but rather is proportional to it. Yet, as the renormalized reaction rate
in $d>2$ is a constant depending on lattice spacing, decay exponents are
predicted by the mean field correctly.
\end{thebibliography}
\end{document}